# Atypical Stable Multipath Routing Strategy in MANET

Tanaya Roy
SDET-Brainware Group of Institutions
398, Ramakrishnapur Road
Kolkata 700124

Sandip Roy
SDET-Brainware Group of Institutions
398, Ramakrishnapur Road
Kolkata 700124

## ABSTRACT
MANET is a collection of mobile nodes operated by battery source with limited energy reservoir. The dynamic topology and absence of pre-existing infrastructure in MANET makes routing technique more thought-provoking. The arbitrary movement of nodes may lead towards more packet drop, routing overhead and end-to-end delay. Moreover power deficiency in nodes affects the packet forwarding ability and thus reduces network lifetime. So a power aware stable routing strategy is in demand in MANET. In this manuscript we have proposed a novel multipath routing strategy that could select multiple stable routes between source and destination during data transmission depending on two factors residual energy and link expiration time (LET) of nodes. Our proposed energy aware stable multipath routing strategy could attain the reliability, load balancing, and bandwidth aggregation in order to increase the network lifetime.

### General Terms
Multipath Routing, Load Balancing

### Keywords
MANET, Stable Route, Link Expiration Time, Energy Consumption

## 1. INTRODUCTION
Here and now wireless technology has gained major importance in industrial environment to comply with requirement pertaining to reliable communication. The potential capability of wireless networks can overcome the inherent difficulties of setting up cable networks. MANET is a self-configuring infrastructure less power constrained wireless network. Frequent failure of nodes driven with limited battery may degrade the communication delay between nodes hampering the QoS of the network. These nodes act as both router and terminal node while forwarding data in the network. Thus based on the network connectivity dynamic routing decision takes place.

The limited bandwidth of wireless link and energy constrained nodes in dynamic network topology of MANET makes routing strategies more complex [1]. Proliferation of MANET have yield several routing strategies that can be categorized in three categories as proactive, reactive and hybrid. Single path routing could not provide much better result due to the unpredictability of the environment and unreliability of the medium. Also, single path routing strategies are not much resilient in handling mobility. Such resilience can be obtained by means of multipath routing.

The primary objective of classical multipath routing is to discover multiple paths between source and destination to countenance load balancing among several nodes, attenuate network traffic and aggregate network bandwidth in order to ameliorate the quality of services (QoS) [2]. So, an efficient multipath routing is required in the context of MANET to proliferate expected performance. In this context we have proposed a multipath routing strategy having trait in selection of neighbor nodes based on link stability in order to upshot the MANET performance in the context of load balancing and data transmission. The link stability is computed based on residual energy and link expiration time of neighbor mobile nodes.

The rest of the manuscript is ordered as follows: section II describes the state of the art review, section III enunciates the proposed routing model assumptions, section IV elaborates the proposed routing methodology, section IV represents the performance evaluation and result analysis and the last one section V is the coda of this manuscript.

## 2. STATE OF THE ART REVIEW
Proliferation in MANET indulges researchers to come up with more efficient routing techniques. Activeness of node and the connectivity duration between nodes play an important role in routing strategy. There are several multipath routing strategies proposed in MANET in attain the quality of services (QoS).

In [3] a multipath routing strategy is proposed based on link stability and energy consumption to improve the network performance. Researchers in [4] have designed an energy efficient routing scheme for MANET considering variable range transmission. Geetha G. and N.J.R. Muniraj has articulated a multipath routing methodology in [5] that provides stability in energy efficient multipath routing to ensure high network lifetime. M. Chakraborty and N. Chaki have proposed a restricted multi-path routing algorithm that can dynamically selects the number of neighboring nodes for transmitting packet from source node to sink node [6]. Authors in [7, 8] have proposed QoS aware multi-path routing for Real-Time Application in WSN.

## 3. PROPOSED ROUTING MODEL ASSUMTION
The most changeable aspect of MANET is mobility of nodes that emphasis on dynamic routing decision. So the selection of link with higher link expiration time yields more stable paths in MANET. On the other hand the selection of nodes with higher residual energy confirms the data transmission reliably. So these two fundamentals play prime importance in selection of a routing strategy.

### 3.1 Residual Energy Computation
Residual Energy (RE) of a node is defined as its remaining energy after data transmission i.e. in terms of initial energy $E_I$ and energy consumed for t bit data transmission $E_C(t)$ as follows:

$RE = E_I - E_C(t)$ …………………….. (1)

## 3.2 Link Expiration Time

Link expiration time (LET) can be calculated by the prediction of future disconnected time between two neighbor nodes in motion. Let us consider two nodes $n_1$ and $n_2$ at $(x_1, y_1)$ and $(x_2, y_2)$ coordinates with uniform transmission radius $r$ and initially they are in transmission range. Let nodes $n_1$ and $n_2$ have speed $v_1$ and $v_2$ along the directions $\theta_1$ and $\theta_2$ respectively. Then link expiration time (LET) between $n_1$ and $n_2$ is defined as follows:

$$LET = (-(ab + cd) + \sqrt{(a^2 + c^2) * r^2 - (ad - cb)^2}) / (a^2 + c^2) \quad \ldots (2)$$

Where $a = v_1 \cos\theta_1 - v_2 \cos\theta_2$ ………. (3)

$b = x_1 - x_2$ …………………… (4)

$c = v_1 \sin\theta_1 - v_2 \sin\theta_2$ ………... (5)

$d = y_1 - y_2$ …………………... (6)

## 4. PROPOSED ROUTING METHODOLOGY

On demand multipath routing strategy has three phases viz. route discovery, route establishment, route maintenance. The process of route discovery searches for multiple paths from considered source to intended destination to transfer the data packets. The route establishment technique set ups the suitable routes among the nodes.

The main objective of our proposed strategy is to launch multiple paths from source to destination through the utilization of powerful nodes with smaller error counts and links with higher link expiration time. Through route maintenance all the established links are cached for a viable timestamp. The algorithm for our proposed routing scheme is excogitated as follows:

**AMR ()**

**Input**: Considered network, Source, Destination

1. Begin

2. int existRoute ← Call Check_Route_Cache()

3. if (existRoute ≠ 0)

3.1 Sent the data packet along the routes enlisted in route cache

4. else

4.1 Call Route_Discovery ()

4.2 Call Route_Establish ()

4.3 Call Route_Maintenace ()

5. End

**Check_Route_Cache ()**

**Input**: Source node, Destination node, timestamp, Route_Cache Table

// each row of Route Cache table consists of id, Source, Destination, Path, Timestamp

1. Begin

2. int existRoute ← 0

3. If (row.source = sourceNode && row.destination = destinationNode && row.timestamp ≤ timestamp)

3.1 Path_id = Path_id ∪ {row.id}

3.2 Increment existRoute by one

4. If (existRoute = 0)

4.1 Return existRoute

5. Else

5.1 Return (existRoute, Path_id)

4. End

**Route_Discovery ()**

**Input**: Source node, Destination node

1. Begin

2. currentNode ← Source node

3. while (currentNode ≠ destinationNode)

3.1 Broadcast Hello message to its neighbor.

// Hello message consists of sender's velocity and direction

3.2 On receiving Hello message each node computes link stability between sender node and itself using the formula below:

Link_stability = $w_1 \times$ (RE) + $w_2 \times$ (LET) ……… (7)

3.3 Sends Reply_Hello_Message to senders.

//Reply_Hello_Message consists of Link_stability between node and sender along the other necessary fields.

3.4 Generates set neighbor Nodes

3.5 currentNode ← node, $\forall$ node $\in$ neighborNode

3.5 End for

4. End

**Route_Establish ()**

**Input**: Set neighbor Nodes for each of the nodes from source to destination nodes

1. Begin

2. currentNode ← sourceNode

3. while (currentNode ≠ destinationNode)

3.1 Evaluates the highest Link_stability among neighborNode (currentNode)

3.2 if (highest Link_stability ≥ $\theta_{highLink\_stability}$)

// $\theta_{highLink\_stability}$ is the lower limiting value for high link

3.2.1 SelectedNode ← {node}, where node is the neighborNode corresponding to the highest Link_stability.

3.3 else if (highest Link_stability ≥ $\theta_{moderateLink\_stability}$)

// $\theta_{moderateLink\_stability}$ is the lower limiting value for moderate link stability

3.3.1 SelectedNode ← {$node_1$, $node_2$}, where $node_1$ and $node_2$ are the neighbor nodes corresponding to the highest Link_stability and second highest Link_stability respectively.

3.4 else

3.4.1 SelectedNode ← {all neighbor nodes}

3.5 currentNode sends RREQ message to each node of selectedNode.

// RREQ includes source node, next node, partial path, Partial_path_stabiliy. If Partial path □ <L1, L2 … Ln>, where Li's are link then Partial_path_stabiliy = ∏ $L_i$ (i = 1 to i).

3.6 If more than one RREQ message is received by any node of selectedNode

3.6.1 It discards the entire RREQ message except the message with maximum Partial_path_stabiliy

3.7 currentNode □ node, ∀ node ∈ selectedNode.
3.8 End While
4. Destination node enlists all the Partial_Path to the routing table along with time stamp from all received RREQ packets.
5. Destination Node sends RREP message along all newly enlisted paths in routing tables.
6. The Source node equally distributes the data packet along the discovered path.

7. Source node sends the data packets along the selected route from the multiple paths.

8. End

**Route_Maintenace ()**

1. Begin

2. If timestamp gets over

    2.1 Delete the entries in route cache

3. End

## 5. PERFORMANCE EVALUATION AND RESULT ANALYSIS

Our proposed multipath routing protocol has been successfully simulated using Matlab 2012b. Let us assume eleven nodes are randomly distributed over 500 m × 500 m terrain area. Maximum communication range of a node has set at 200 m. Initial energy of each node is assumed as 5J. MICA2 energy model is used for transmitting and receiving signals [9].

We have assumed the amount of energy required for receiving and transmitting signals are 0.234 μJ/bit and 0.312 μJ/bit respectively. Let us assume each node is moving with constant velocity 10 m/s along with given direction enlisted in Table 1. As shown in Fig. 1, there are three feasible paths from source node 'A' to destination node 'K'.

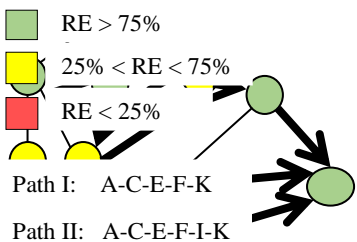

Fig. 1. Our proposed Multipath Routing Strategy

**Table 1. Residual Energy of All Nodes along with speed and moving direction**

| Node ID | Residual Energy (%) | Speed (m/s) | Moving Direction (In degree) |
|---------|---------------------|-------------|------------------------------|
| A | 90 | | 49 |
| B | 85 | | 35 |
| C | 78 | | 83 |
| D | 70 | | 65 |
| E | 68 | | 48 |
| F | 85 | 10 m/s | 40 |
| G | 50 | | 50 |
| H | 20 | | 34 |
| I | 60 | | 69 |
| J | 78 | | 45 |
| K | 95 | | 75 |

### 5.1 Performance Analysis

**Performance metrics:** The performance of our algorithm is analyzed according to the following metrics:

**Packet Delivery Ratio (PDR):** This is the ratio of the number of data packet received and the number of packet send.

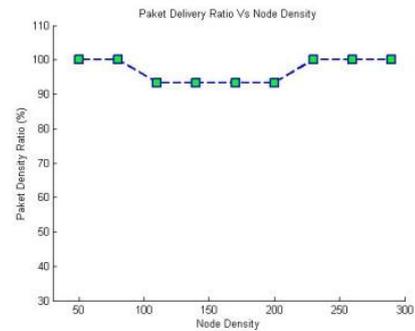

**Fig. 2. Packet Delivery Ratio Vs Node Density (Fixed Mobility 10 m/s)**

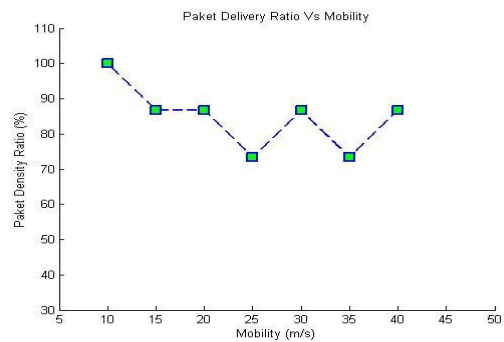

**Fig. 3. Packet Delivery Ratio Vs Mobility (Fixed Node density as 100 nodes in 500 m × 500 m area)**

**Average Energy Consumption Rate:** Average Energy Consumption Rate is defined as the percentage measure of energy consumption.

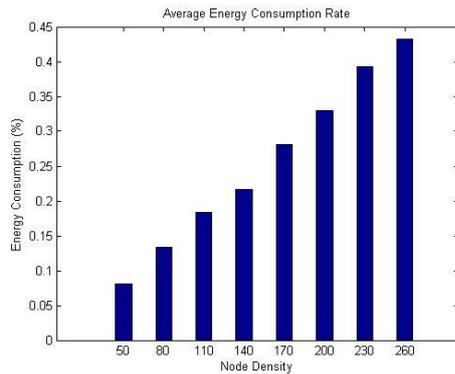

**Fig. 4. Energy Consumption Rate Vs Node Density**

## 6. CONCLUSION AND FUTURE WORK

This context a stable multipath routing strategy is discussed. Using our proposed routing method, a node is selected based upon its' current energy and corresponding link expiration time. Our simulation results show that average 90% successful packet delivery in different node density and mobility conditions. In future we would like to implement a data delivery method along the multiple path simultaneously.